\documentclass[review]{elsarticle}

\usepackage{hyperref}
\usepackage{amsmath}

\usepackage{graphicx}
\usepackage[english]{babel}
\usepackage{dirtytalk}
\journal{New Astronomy}
\usepackage{natbib}
\usepackage{subfiles}

\begin{document}

\begin{frontmatter}

\title{Re-evaluation of $\Omega_k$ of the normalised Friedmann-Lema\^{\i}tre-Robertson-Walker model: Implications for Hubble constant determinations}


\author{Ahmet M. \"O\/zta\c{s}}
\address{Physics Engineering, Hacettepe University, TR-06800 Ankara, Turkey}

\author{Michael L. Smith*}
\address{Ume\aa \hskip1mm University, SE-907 38 Ume\aa \hskip1mm Sweden}


\ead{mlsmith55@gmail.com}


\begin{abstract}
The description of spacetime is an fundamental problem of cosmology. We explain why the current assignments of spacetime geometries for $\Omega_k$ of the Friedmann-Lema\^{\i}tre-Robertson-Walker (FLRW) model are probably incorrect and suggest more useful descriptions. We show that $\Omega_k$ represents not only curvature but the influence of matter density on the extent of spacetime between massive objects. Recent analyses of supernovae type Ia (SNe Ia) and HII/GEHR data with the FLRW model present the best fits with a small value for $\Omega_m$ and a large $\Omega_k$. These results are consistent with our Universe exhibiting sparse matter density and quasi-Euclidean geometry and the small $\Omega_m$ value agrees with Big Bang nucleosynthesis calculations. We suggest the geometry of our current Universe is better described by a value for $\Omega_k\approx$1 rather than 0. As an example we extend the FLRW model towards the Big Bang and discover a simple explanation of how matter creation developed into the currently geometrically flat Universe with sparse, homogeneous, isotropic matter and energy distributions. Assigning  $\Omega_k\approx1$ to describe quasi-Euclidean spacetime geometry is also useful for estimating $H_0$ and should help resolve the \say{tension} surrounding current estimates by different investigators.
\end{abstract}

\begin{keyword}
\texttt (cosmology:) cosmological parameters\sep (cosmology:) early universe \sep cosmology: theory

\end{keyword}

\end{frontmatter}


\section{Introduction}
A central problem of cosmology is explaining the interplay of the three fundamentals; matter, energy and spacetime. Though we think we understand spacetime there is no consensus about what constitutes spacetime, how it got here or how it functions. Some people believe spacetime evolved entirely from the Big Bang (BB), some not. Some think the spacetime between galaxies is itself increasing with time and the rate now accelerating, some not. The broad applicability of the Friedmann-Lema\^{\i}tre-Robertson-Walker (FLRW) model is also often used to bolster the argument that spacetime itself is expanding but the model is really noncommittal. 

A major problem is explaining why our Universe seems geometrically flat, homogeneous and isotropic in matter and light when assayed to great distance. The notion that our Universe is really absolutely flat is now questioned \cite{Witzemann2018,Valentino2020}. Einstein argued that a universe containing matter, spacetime geometry must always be roughly elliptical, with or without the cosmological constant $\Lambda$, and is currently quasi-Euclidean \cite{Einstein1952}. The results from observations of supernovae type Ia (SNe Ia) emissions and some interpretations of the cosmic microwave background (CMB) indicate to some observers quasi-Euclidean and endless spacetime \cite{Riess2001,Abbott2019,Collaboration2018a}. Re-examination of the CMB signals suggests to others that spacetime geometry is elliptically curved and the Universe closed; the disagreement is considered a \textit{crisis} \cite{Valentino2020}.

The properties of the term for spacetime, $K$ of the second Friedmann relationship, are followed by many the past three decades \cite{Carroll1992}. Since $K$ is but the constant of integration some admit we really don't know what it represents. Krizek and Somer \cite{Krizek2016} have recently shown that some of the current assumptions about the geometries assigned to values of $K$ are simply wrong. An example is the impossible situation of hyperbolic spacetime curvature which is only possible in the presence of negative matter density. The problems with such assignments coupled with the straight-forward argument of Einstein has lead us to rethink the current method used for evaluation of critical SNe Ia data and the interpretation of normalised term for spacetime, $\Omega_k$. 

There is also ongoing disagreement, termed \textit{tension}, over the value of the local Hubble constant $H_0$ \cite{Rameez2019, Rejkuba2020,Valentino2020c}. Part of the blame for this and the crisis over spacetime geometry is the common notion that our universe exhibiting flat, or nearly flat geometry, demands an $\Omega_k\approx0$. ($\Omega_k$ and $\Omega_m$ as described in subsection \ref{assignments}.) Using proper regression analyses with several public collections of SNe Ia data we have shown that the FLRW cosmology without the normalised cosmological constant, $\Omega_{\Lambda}$, is the preferred model with values of $\Omega_k$ approaching 1 and with $H_0$ less than 70 km s$^{-1}$ Mpc$^{-1}$ where Mpc is megaparsec \cite{Oeztas2008,Oeztas2014,Oeztas2015}. We have more recently found an $\Omega_k$ approaching 1 from evaluating HII and giant extragalactic HII regions (GEHR) data \cite{Smith2020}. Results from the Freedman group using emissions from the tip of the the red giant branch \cite{Freedman2019a} and detection of a bineutron star collision by the LIGO consortium \cite{Collaboration2017} both point to a $H_0$ of about 70 km s$^{-1}$ Mpc$^{-1}$. Our reinterpretation of the $\Omega_k$ assignments supports a nearly flat and slightly elliptical Universe geometry for the current epoch, consistent with a new evaluation \cite{Valentino2020}. We shall show that one is not forced to find $\Omega_k\approx0$ to explain a quasi-Euclidean Universe geometry and a value for $H_0\leq70$ is very possible. 

To explain why the geometry of our Universe seems flat with homogeneous, isotropic matter and energy distributions at large scales, early investigators \cite{Guth1981,Linde2007} argue the early Universe suffered exponential expansion from the BB in an inflationary process. One version is now the \say{chaotic inflation} model \cite{Linde1986,Kallosh2014} and another version has become the \say{multiverse} hypothesis \cite{Linde2015}. Despite, or perhaps because of these many variations, the inflationary universe explanation is currently popular as evidenced by the many reviews \cite{Gonzalez2020}. The key mechanism of inflation involves some form of energy storage in the early Universe followed by rapid release - which may or may not be related to the dark energy supposedly responsible for our accelerating Universe expansion. Some people however, strongly object to this theory \cite{Ijjas2014,Martin2015}. An outstanding problem is how did it all end, that is, why and by what mechanism stopped inflation, or is inflation eternal \cite{Guth2007}. Inflationary theory has developed many variations, any one which fits hand-selected observations and there are so many versions that it is impossible to fairly test all \cite{Ijjas2013}.

We remind the reader that the properties assigned to $K$ are only suggestions and subject to revision when confronted with data. Here we extend the meaning of $K$, in the absence of a significant $\Lambda$, to represent everything else in the universe, other than matter, which means primarily spacetime. As an application of our finding we then employ the FLRW model to trace the normalised matter parameters, $\Omega_{rm}+\Omega_m$, where $\Omega_{rm}$ represents the input of relativistic matter, and $\Omega_k$ values during particle creation in an early universe. Because of the interdependence of the matter and spacetime parameters we find the value for $\Omega_k$ fluctuated wildly in the early Universe and the behaviour more interesting than previously allowed. We trace the values of $\Omega_{rm}+\Omega_m$ and $\Omega_k$ after particle creation and show why these are important to understand the current dilute matter and energy distributions and our nearly, but not quite flat, Universe geometry. We think our introductory view allows more thorough study of inflationary theory and helps resolve the current tension about the $H_0$ value.

\section{Theory}
We conservatively consider only energy, confirmed matter types and 4-dimensional spacetime after the Planck-inflationary epoch(s); the standard model of particle physics without the necessities of $\Lambda$ or supersymmetry.

\subsection{The cosmological constant problem}
The cosmological constant has been a fashionable topic the past two decades. Since the initial announcements supporting a positive $\Lambda$ there are many supportive opinions and reviews \cite{Sahni2000,Riess2019}. Here are the many reasons we do not incorporate $\Lambda$ in our analysis.

The energy associated with $\Lambda$ is volume dependent and tiny so the total $\Lambda$ energy would have been minuscule in the early Universe and  the effect of $\Lambda$ is only expected to become noticeable at $\approx3.5$ billion years \cite{Barnes2018}. The value of $\Omega_{\Lambda}$ was investigated, beginning with a value currently favoured by astronomers ($\Omega_{\Lambda}\approx0.7$), then tracing this towards the BB. It was discovered that $\Omega_{\Lambda}$ must become incredibly large, $\gg1$, with increasing lookback demanding either $\Omega_k\ll-1$ or $\Omega_m<0$ and both requirements lack reality \cite{Oeztas2018}. Another way of putting this is when $\Omega_{\Lambda}\gg1$ as required by normalisation for accelerating Universe expansion, $\Omega_m+\Omega_k$ must become $\ll0$. Solutions with $\Omega_k<0$ denoting hyperbolic curvature or $\Omega_m<0$ demanding negative matter density are unrealistic and of mathematical interest only \cite{Krizek2016}.

The SNe Ia data are subject to random and systematic errors; the signals are source dependent and evolve over time \cite{Foley2019a, Kang2020} and the random errors of distance measurements, as $D_L$, are huge \cite{Oeztas2015}. According to some investigators when the local anisotropy is taken into account, evidence for accelerating Universe expansion disappears along with the need for dark energy \cite{Colin2019a,Colin2019b}. Being the major force in the Universe dark energy should be testable but a sensitive laboratory experiment failed to detect any such force \cite{Sabulsky2019}. The value for $\Lambda$ suffers a difference from the required vacuum energy by a factor of $10^{30}$ or more, the \textit{vacuum catastrophe} as noted by some early proponents \cite{Carroll2001,Frieman2008}, but this has never been properly explained away. Some variations of the dark energy hypothesis and some criticisms have been addressed \cite{Brax2017,Rubin2020}. Problems with $\Lambda$ have even been noted in the popular press eliciting rebuttal from major proponents \cite{Riess2018}.

\subsection{SNe Ia data analysis}
\subsubsection{The SNe Ia data analysis problem}
The case for a real $\Lambda$ and a flat universe geometry with $\Omega_k=0$ began in the last millennia with three publications. The first was a review article which made the idea of a real $\Lambda$ seem very interesting \cite{Carroll1992}. Later that decade two groups published results from investigations of emissions from distant SNe Ia which were interpreted as support for a real $\Omega_{\Lambda}$ and hence a real $\Lambda$ with $\Omega_k=0$ describing flat universe geometry \cite{Riess1998,Perlmutter1999}. Both groups analysed the SNe Ia data using not the luminosity distances, $D_L$, to SNe Ia but values termed \textit{distance magnitude}, often simply \textit{mag}, based on the relationship
\begin{equation}
mag=5\hskip1mm log(D_L)+25.
\label{logequa}
\end{equation}
Neither group correlated $D_L$ with expansion factors or recession velocities, for the independent variables, but the observed redshifts, $z$. The incorrect use of \textit{mag} and redshift rather than $D_L$ \textit{vs.} the expansion factor (or recession velocity) has been noted \cite{Smith2020}. That article explains in detail the many faults of this common method and the rationale for reanalysis of SNe and HII/GEHR data. Using log-transformed data, such as equation (\ref{logequa}), often leads to erroneous findings \cite{Feng2014,Packard2010}. 

Our results of correlating $D_L$ \textit{vs.} the expansion factor using the SNe Ia data first published by the High-Z Supernova Search Team \cite{Riess1998}, are shown in Table \ref{smalldataset}. The value of $H_0$ for both determinations is under 63 km s$^{-1}$ Mpc$^{-1}$. Similar results from analyses of more recent, larger sets of SNe Ia data have been reported \cite{Oeztas2008,Oeztas2014,Oeztas2015}.

\begin{table}[htb]
\caption{Analysis of $D_L$ \textit{vs.} expansion factor with 37 observations and the local group (1,0) using the robust regression technique \cite{Smith2020} For the reduced $\chi^2$ where N is the number of data pairs (38) and FP the number of free parameters, the better fitting model exhibits lower reduced $\chi^2$. The best fit curve is presented in Figure \ref{fig:DistanceData}}.
\label{smalldataset} 
\begin{tabular}{lcccc}
\hline
\smallskip
		Model & $\Omega_m$ & $\Omega_k$ or            & $\chi^2$/(N-FP) &       $H_0 $                \\
		      &            &  $\Omega_{\Lambda}$      &                  &     km s$^{-1}$ Mpc$^{-1}$ \\
		\hline
		\vspace{4mm}
		$\Omega_m+\Omega_{\Lambda}=1$ & 0.37$\pm$0.13 & 0.63$\pm$0.13  & 1.50  & 62.9                   \\
	    $\Omega_m+\Omega_k=1$         & 0.01$\pm$0.23 & 0.99$\pm$0.23  & 1.44  &  62.6                   \\
\hline
\end{tabular}
\end{table}

\subsubsection{The Hubble correlation}
Real distance and recession velocity data should be analysed in a manner consistent with that originally used by Hubble to calculate $H_0$ \cite{Hubble1929}. Use of that analytical technique, with tip of the red giant branch data, yielded a value of $\approx70$ km s$^{-1}$ Mpc$^{-1}$ \cite{Freedman2019a} similar to other independently determined values \cite{Collaboration2017}. The FLRW model may be used, however, and the analysis conforms with the requirements of physics when $D_L$ and expansion factor data are correlated.

\subsection{Current spacetime curvature assignments}
\label{assignments}
The development of our Universe on the largest scale can be approximated by the two Friedmann expressions relating the expansion factor, \textit{a}\hskip0.3mm, with matter density $\rho$, pressure $p$, $c$ lightspeed and $\ddot{a}$ acceleration  with the first equation as
\begin{equation}
\label{Fried1}
\frac{\ddot{a}}{a}=-\frac{4 \pi G}{3}\left(\rho+\frac{3p}{c^2}\right) \,,
\end{equation}
and the second equation being
\begin{equation}
\label{Fried2}
\left(\frac{\dot{a}}{a}\right)^2=\frac{8\pi G}{3}\rho - \frac{Kc^2}{a^2}=H^2 \,,
\end{equation}
with values of $H, \rho, p, K$ and $a$, all greater than zero for our Universe with $\Lambda=0$ during the first epochs. This $K$ is usually assigned a value describing constant spacetime curvature in the presence of conserved matter density as
\begin{align}
\text{when} \hskip1mm K
\begin{cases}
& 0<  \hskip6mm   \text{elliptical,}\\
&=0   \hskip7mm      \text{Euclidean,} \\
&<0  \hskip7mm   \text{hyperbolic.}
\label{Kgeometry1}
\end{cases}
\end{align}
The assignments are sometimes described with different terms; $K>0$, closed; $K=0$, flat; $K<0$, open and Einstein considered nearly flat to be quasi-Euclidean. We remind the reader that $K$ is unit-less, so gives no hint of meaning and the properties above have been assigned and never really verified. It may be that $K$ simply indicates that portion of the universe which is not matter rather than only curvature.

More detailed descriptions of the various meanings and conditions commonly assigned to $K$ are presented in detail \cite{Piattella2018} and with more geometry \cite{Wesson2002}. The case where $K=0$ is a special situation and has problems due to over-interpretation \cite{Krizek2016}. Also note our Universe has never been observed to display hyperbolic geometry; geometric spacetime geometry is unrealistic in the presence of ordinary matter.

One often considers the Universe of general relativity suffering spherical expansion as described by the Robertson-Walker metric 
\begin{equation}
\label{10}
ds^2=-dt^2+a^2(t)\left[\frac{dr^2}{1-kr^2}+r^2d\theta^2+r^2\sin^2\theta d\phi^2\right] \,,
\end{equation}
where $r, \hskip1mm \theta$ and $\phi$ are the usual designations associated with spherical geometry. For practical reasons the various terms of this model are normalised according to the following conditions
\begin{align}
\rho_s= & \frac{\rho_{i}}{\rho_c},  & \Omega_{rm},\Omega_m= & \frac{8\pi G}{3 H^2}\rho_s>0,  &  \Omega_r=&\frac{8\pi G}{3 H^2}\rho_r,   &  \Omega_k= & -\frac{Kc^2}{a^2 H^2}\,,
\label{normals}
\end{align}
where $\rho_i$ represents the densities of the various massive particle forms; small relativistic particles such as neutrinos, with tiny mass, are considered differently from massive particles. Above $\rho_c$ is the critical particle density and the ratio of $\rho_i$ to $\rho_c$ is important because $\frac{\rho_i}{\rho_c}>1$ encourages eventual universe collapse. Here $\rho_r$ is radiation density, $\Omega_{r}$ the normalised radiation density with $\Omega_{rm}$ and $\Omega_{m}$ the normalised matter densities suffering relativistic and non-relativistic velocities, respectively. During the early Universe, with a significant influence of relativistic massive particles, the value for $\Omega_{rm}$ cannot be treated in the same manner as non-relativistic, baryonic matter, $\Omega_m$. All types of pure energy density, from radio to hard $\gamma$-rays including exotic entities are grouped together as $\rho_r$ and $\Omega_r$ and we remind the reader that pure energy has never been shown or observed to exert gravity.

When describing our current epoch we employ the normalised relationship without $\Omega_{\Lambda}$, for the many reasons given above, as 
\begin{equation}
\Omega_m+\Omega_k+\Omega_r = 1\,,
\label{normal}
\end{equation}
where $\Omega_{rm}$, consisting mainly of neutrinos, exerting a negligible effect during the current epoch, is also ignored. (While the neutrino population is large the gravitational effect is negligible.) 

We think some presumptions around the nature of $\Omega_k$ are wrong. We remind the reader that there is no firm rule demanding the properties often assigned to $\Omega_k$ and $K$. We cannot also presume a direct correspondence between $K$ and $\Omega_k$, because the very act of normalisation automatically considers matter density. Nowadays many astrophysicists claim the current $\Omega_m\approx$ 0.25 to 0.30 with $\Omega_{\Lambda}\approx0.75$ and an $\Omega_k\approx 0$ means a geometrically flat spacetime. Flat spacetime geometry cannot possibly be true with such a large value for $\Omega_m$, with or without $\Omega_{\Lambda}$ \cite{Einstein1952}. The assignment of Euclidean geometry to $K=0$ (and $\Omega_k=0$) comes from later investigators and was not considered by Friedmann. For these reasons it is best to be wary of evaluating $K$ or $\Omega_k$ at 0 \cite{Krizek2016}. It is not always obvious if reports refer to spacetime curvature in terms of $K$ or $\Omega_k$, from here on we are interested in $\Omega_k$ and the reader may trace our argument to $K$ for themselves.

\subsection{Radiation and gravitational attraction}
Since there is no experimental evidence nor robust observational report that the photon has measurable mass or exerts gravitational attraction, the gravitational effect of $\Omega_r$ in reality is negligible. One should realise that a photon following a geodesic is evidence that matter can bend spacetime but is not evidence that light can bend spacetime. (There is a study attempting to measure the gravitational pull of the photon which reports difficulties with large thermal effects and systematic error \cite{Rancourt2015}). One may use terms for radiation in the Einstein stress-energy tensor but this is not evidence that massless particles elicit gravitational attraction as we have pointed out \cite{Smith2017}.

 \subsection{Suggested spacetime geometry curvature properties}
Spacetime geometry as $\Omega_k$ in the presence of matter cannot be either hyperbolic nor Euclidean but only elliptical. Here we propose these spacetime geometries associated with $\Omega_k$ for our Universe as
 \begin{align}
\text{when} \hskip1mm \text{$\Omega_k$}
\begin{cases}
&= 1,  \hskip10mm   \text{flat, Euclidean}\\
& <1,   \hskip10mm      \text{elliptical}
\label{Kgeometry2}
\end{cases}
\end{align}
and do not consider the geometries of other possible universe types.

Here are some reasons for our proposal. Consider the value for the limiting case before particle creation when $\Omega_k=1$ and $\Omega_m=0$. In the early universe $\Omega_m\approx 0$ leaving $\Omega_k\approx1$ as per equation (\ref{normal}). With but little gravity our Universe geometry would have been quasi-Euclidean, $H$ very large, allowing energy to travel anywhere probably at near lightspeed. On the other hand in a system cramped with matter, such as a black hole with $\Omega_m\geq1$, $\Omega_k$ will be $\leq0$, with little spacetime between particles and this small world must exhibit highly elliptical geometry.
 
There are three important properties of our Universe and $\Omega_k$, $\Omega_m$ consistent with our proposal. First, it is a long distance between stars and between galaxies; spacetime is nearly devoid of matter. Second, spacetime appears quasi-Euclidean for an observer distant from a massive object. Third, an $\Omega_k$ near 1 and a small $\Omega_m$, as presented in Table \ref{smalldataset}, indicates great spacetime between objects and slightly elliptical universe geometry. There is also an important non-observation - there are no reports of hyperbolic spacetime curvature.

Extending this reasoning further, we think the value of $\Omega_k$ indicates the relative importance of spacetime itself as well as geometry. Normalisation really means just that - the values for $\Omega_{rm}$, $\Omega_m$ together with $\Omega_k$ reflect the average matter densities and the average spacetime between massive objects as well as spacetime geometry.

 \section{Results}
\label{results}
As an example of usefulness of this concept we apply the new assignments of $\Omega_k$ geometry to the first few seconds after the BB. We first evaluate the FLRW model with increasing values of $\Omega_k$ as a function of \textit{a} with $\Omega_m$ approaching 0 with constant $H$ and evaluate $\Omega_k$ again as $\Omega_r$ disappears. We next apply our new understanding of $\Omega_k$ to the problem of the geometry of the early Universe and to the observations of homogeneity and isotropy of matter and energy distributions in our Universe.

\subsection{The FLRW model applied to the current and early universe}
Equation (\ref{supernova}) presents a relationship from the FLRW conditions which can be used for testing cosmological models of the current epoch with SNe,  $\gamma$-ray burst and HII/GEHR emissions. The values for $D_L$ represent standard candle distances while the associated redshifts are used to calculate the expansion factor, \textit{a} = 1/(1+\textit{z}). For tests using the FLRW model without $\Omega_{\Lambda}$ the relationship correlating distance with recession velocity is
\begin{equation}\label{supernova}
D_L=\frac{c}{a H_0\sqrt{\lvert\Omega_k\rvert}}\text{sinn}\Bigg[\sqrt{\lvert\Omega_k\rvert}\int_{a_1}^1\frac{da}{a\sqrt{\frac{\Omega_r}{a^2}+\frac{\Omega_m}{a}+\Omega_k}}\Bigg].
\end{equation}
Here sinn is \textit{sin} when $\Omega_k<0$ or \textit{sinh} when $\Omega_k>0$. As instructed by Carroll {\em et al.} and by Riess {\em et al.} \cite{Carroll1992,Riess1998} equations (\ref{supernova}) and (\ref{earuni}) should be evaluated with $\sqrt{\vert\Omega_k\vert}=1$ when presuming $\Omega_k=0$ for a flat universe geometry, which is an obvious inconsistency. We rather suggest that assigning the value of $\Omega_k=1$ for Euclidean geometry means just that and should be used in equations (\ref{supernova}) and (\ref{earuni}) when presuming flat or quasi-Euclidean spacetime geometry. In the case where $\Omega_k=0$ it really means that $\frac{\rho}{\rho_c}=1$ and equations (\ref{supernova}) and (\ref{earuni}) cannot be evaluated. (The situation where $K=0$ was never addressed by Friedmann \cite{Krizek2016}.) When $\Omega_r$ is included in equation (\ref{supernova}) during problem presentation it is \textit{always} dropped during analysis of astronomical data, the reason given as the effect of radiation during our current epoch simply being too small to affect measurements and astronomical observations. 

Allowing the gravitational effect of radiation as unimportant leaves us the simpler situation of
\begin{equation}\label{simpUniverse}
\Omega_m+\Omega_k=1 \,,
\end{equation}
with only two normalised parameters describing the Universe. For the condition where $\Omega_m=0$ applies for a universe without matter, this relationship must allow $\Omega_k=1$ as describing an empty universe with Euclidean geometry while presuming a real $\Omega_r$ but no matter and is only useful for speculation (Appendix B).

\begin{figure}
	\includegraphics[width=\columnwidth]{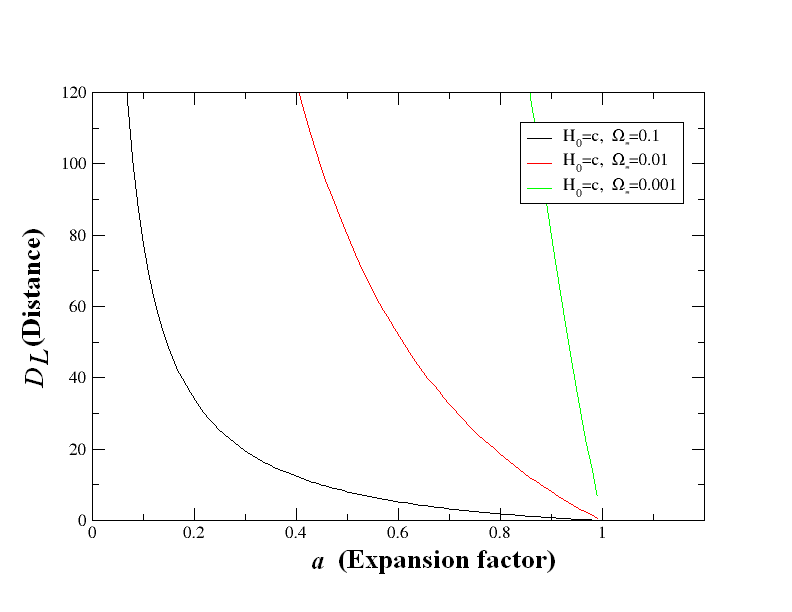}
    \caption{$D_L$ \textit{vs.} \textit{a} with $H_0=c=1$ for a 
    universe with $\Omega_m$ from 0.1 to 0.001, left to right, with corresponding 
    values of $\Omega_k$, calculated with equation (\ref{supernova}) and $\Omega_r= 0$.}
    \label{fig:Fig1}
\end{figure}

If we allow $\Omega_m=0$ and $\Omega_k=1$ to describe our early Universe just after the BB, expression (\ref{simpUniverse}) collapses to 1 so is not terribly informative. We think $\Omega_m\approx0$ a reasonable assumption during that epoch; the Universe before massive particle creation consisted primarily of high energy but not matter. To get around this problem we evaluate a universe with $\Omega_m$ decreasing from 0.1 to 0.001 using equation (\ref{supernova}) without the $\Omega_r$ term and results presented as Figure \ref{fig:Fig1}. The distance allowed for light travel increases drastically with decreasing $\Omega_m$. That is, $D_L$ approaches extreme dependence on the expansion factor with decreasing matter density but at \textit{a} of 1 most any distance is allowed. This property of spacetime, as expressed in allowed light flight, is expected as elliptical geometry relaxes towards quasi-Euclidean. An $\Omega_k$ of 1 is consistent with a flat universe devoid of gravity with radiation everywhere in a universe without bounds.

To consider $D_L$ for a universe before particle creation we must drop $\Omega_m$ and consider $\Omega_r$. (For massive particles $G$ obviously applies, but for a system consisting of only high energy species, such as photons, Newton's $G$ is irrelevant.) $D_L$ becomes highly dependent on the expansion factor, with the parameters $\Omega_r$ and $\Omega_k$ as
\begin{equation}
\label{earuni}
D_L=\frac{c}{a H_0\sqrt{|\Omega_k|}}\text{sinn}\Bigg[\sqrt{|\Omega_k|}\int_{a_1}^1 \frac{da}{a\sqrt{\frac{\Omega_r}{a^2}+\Omega_k}}\Bigg].
\end{equation}
A similar situation to Figure \ref{fig:Fig1} is observed in Figure \ref{fig:Fig2} (Appendix B) when the expression (\ref{earuni}) is used to estimate $D_L$ as $\Omega_r$ approaches 0, supposing that radiation does have some gravitational influence (which has never been observed). As the influence of radiation wanes the values for allowed light travel is again very dependent on the expansion factor. In other words, in the early universe devoid of gravity and $\Omega_k=1$, travel by light is unbounded in all directions.

\subsection{Spacetime response to particle genesis}
\label{spacetimeResponse}
As an example of how our new realisation can be applied we use $\Omega_m$ and $\Omega_k$ as tools exploring the characteristics of the Universe just after the BB. We present the results of our calculations in broad terms concentrating on the nature of spacetime, as $\Omega_k$, correlating with the current understanding of particle creations. We only calculate initial particle creations as rapid auto-catalytic events, but other mechanisms were probably involved.  

It is now thought the first epoch was dominated by only high energy species including some exotic particles followed by quark-hadron matter creations a few moments later \cite{Bonometto2016}. Three phase transitions are firmly established; first the electroweak (EW) transition at 100 to 200 GeV, next the quantum chromodynamic (QCD) transition at 150 to 180 MeV, third the epoch of stable electron-positron creation (EP) at about 170 keV \cite{Fu2015,Cyburt2016}. During the energy dominated epoch, matter was completely absent with $\Omega_{rm} +\Omega_m=0$ therefore $\Omega_k=1$. The energy content, if considered as matter equivalent, would probably have exhibited a density, $\rho_r > \rho_c$ in a small volume, but being only high-energy radiation $\rho_r$ exhibits no gravity and $\Omega_r$ was immaterial with respect to spacetime curvature. Therefore $\Omega_k=1$ during this epoch means expansion at near lightspeed in a geometrically flat Universe.  

\subsection{Spacetime geometry, matter and energy dispersal}
The EW transition was quickly followed by the QCD transition as particle creation began. Considering the conditions for equation (\ref{supernova}), $\Omega_{rm}$ and $\Omega_m$ were still tiny, reflecting few particles, and because these parameters are normalised $\Omega_k$ remains $\approx1$. At that instant the positive value for $\Omega_k$ represented not only slight elliptical curvature but also a large spacetime contribution to the Universe in which massive particles were still rare. 

The creation of the first particles perturbed the dense energy ensemble spurring further particle creations in a catalytic manner. As the Universe continued to cool below the EW and QCD transitions, more massive particles were rapidly created close to the first initiating particles. The local values for matter density, $\rho_s$, $\Omega_{rm}$ and $\Omega_m$ rapidly and drastically increased along with the compensatory decrease of $\Omega_k$. The creation processes were probably drastically prolonged due to both particle-antiparticle annihilation and reduction of spin degrees of freedom number from around 60 to about 15 shortly followed by the quark-hadron transitions requiring significant entropy input \cite{Bonometto2016}. Spin redistribution would have prolonged the entropy input and allowed relatively constant temperature.

We model the creation of massive particles where $n\gamma$ represents the large number of high-energy species just after the BB with the forward rate described by
\begin{equation}
n\gamma\hskip1mm +\hskip1mm initiation\hskip1mm particles \hskip1mm \overset{k_1}{\Rightarrow} \hskip1mm massive\hskip1mm particles
\end{equation}
and the reverse reaction described by
\begin{equation}
massive\hskip1mm particles \hskip1mm  \overset{k_{-1}}{\Rightarrow}\hskip1mm n\gamma
\end{equation}
where the forward rate constant, $k_1$, is a slightly larger than the reverse, $k_{-1}$ and the term \textit{massive particles} includes both relativistic and non-relativistic matter. The increase in matter concentration, primarily quarks and hadrons $[m]$, with respect to time, $t$, can be written as
\begin{equation}
\frac{[dm]}{dt}=k_{1}[\gamma][m]-[m]^2
\label{eq:overrate}
\end{equation}
where $[\gamma]$ is the local, high-energy species concentration. For simplicity we model only two species, high-energy $\gamma$ species and massive particles $[m]$.

Since massive particles were initially rare, the second-order annihilation would have been negligible; the above is simplified by ignoring the second term. The rate of $[\gamma]$ disappearance can be approximated by
\begin{equation}
\frac{-d[\gamma]}{dt}=k_{1}[\gamma][m].
\label{eq:simplerate}
\end{equation}
If we now also use estimates for the initial concentrations of $m$ and $\gamma$ as $[m_i]$ and $[\gamma_i]$ we can write the reaction(s) as an auto-catalytic event by generalizing with an overall rate constant, $\mathbf{k}$ not necessarily the same value as $k_1$, with 
\begin{equation}
\frac{-d[\gamma]}{dt}=\mathbf{k}\Bigg([\gamma][\gamma_i]+[\gamma_i][m_i] -[\gamma]^2\Bigg).
\label{autocatalytic}
\end{equation}
Rearranging equation \ref{autocatalytic} and integrating both sides we have
\begin{equation}
-\mathbf{k} \int dt =\int\frac{d[\gamma]}{[\gamma][\gamma_i]+[\gamma_i][m_i] -[\gamma]^2}.
\label{integrateautocat}
\end{equation}
The forward reaction creating matter is described by the relationships 
\begin{equation}
[m] = \frac{[m_i]+[\gamma_i]}{1+\frac{[\gamma_i]}{[m_i]}\text{e}^{-([m_i]+[\gamma_i])\mathbf{k}t}}\approx\frac{[\gamma_i]}{1+\frac{[\gamma_i]}{[m_i]}\text{e}^{-[\gamma_i]\mathbf{k}t}} \,,
\end{equation}
where the approximation is made on the right-hand side, for the condition where $[\gamma_i]\gg[m_i]$. This is simply a general equation for the auto-catalytic creation of most any quark, hadron or lepton which may have been initiated by most any particle type. Here we only consider the overall reaction(s) by high-energy species creating massive particles as an auto-catalytic process occurring with a nearly constant concentration of high-energy species where $[\gamma_i]$ is at least $200> [m_i]$. Some exotic matter would have been created more rapidly after the BB increasing the initiating particle concentration but we ignore this effect on overall $[m]$ to keep our model simple.

Because quark types probably had different formation rates a proper description of quark creations would be a linear combination of several equations with different rate constants. Since the critical temperatures for both quark and hadron creations are similar but not identical, the rate equations describing quark-hadron creations should also be included in a more detailed description. All of this means the time available for quark-hadron creations from high energy species would have been prolonged. 

Stable leptons would have also been created to preserve electroneutrality, but later at lower energies only slightly affecting gravitational attraction. While the Universe was expanding there was simultaneous dilution of both $[m]$ and $[\gamma_i]$. A simple version of all these events is modelled by 
\begin{equation}
[m] = \Bigg(\frac{[\gamma_i]}{1+\frac{[\gamma_i]}{[m_i]}\text{e}^{-[\gamma_i]\mathbf{k}t}}\Bigg) \div  \bigg(2\pi^2 r^3\bigg) \,,
\label{matterdrag}
\end{equation}
where the denominator represents the time-dependent dilution of this process by the growing hypersphere of the early Universe.

Using equation (\ref{matterdrag}) we illustrate the rapid increase of $\Omega_{rm}$+$\Omega_m$ as these species are being simultaneously diluted in an expanding universe in Figure \ref{fig:Figure2}. We allow $\Omega_{rm}$+$\Omega_m=[m]$ with $[\gamma_i]$ being about $10^7$ times that of $[m_i]$ and presume a \textbf{k} of 0.01 as the forward rate constant describing the inefficient creation of matter in an energetic, matter-antimatter-photon-gluon ensemble. The initial lag, when $\Omega_m=0$ and $\Omega_k=1$, represents that time when the Universe consisted only of high-energy species too energetic to create stable matter, exhibiting little gravitational attraction, hence the Universe just after the BB was geometrically quasi-Euclidean.

After this short period, the catalytic particles initiated rapid increases of matter density and $\Omega_{rm}$+$\Omega_m$ followed by both concentrations declining with time. The somewhat gentle nature of the curve increases in Figure \ref{fig:Figure2} may have been even more gradual than illustrated, since $k_{-1}$ was on the order of $k_1$ at high energies. The increase of $\Omega_{rm}$+$\Omega_m$ may have also exhibited a flattened maximum if the range of various overall rate constants, \textbf{k}, was large. Another reason for a complicated increase of $\Omega_{rm}$+$\Omega_m$ is the quark-hadron transitions which occurred quickly after quark creations, also affecting matter characteristics and $\Omega_{rm}$+$\Omega_m$ concentrations. Very general descriptions of particle creation with concurrent dilution into the expanding Universe, as reflected by the probable developments of $\Omega_{rm}$+$\Omega_m$ and $\Omega_k$, are illustrated by Figures \ref{fig:Figure2} and \ref{fig:Figure3}.

\subsection{FLRW model reflecting matter creation and dispersal}
\begin{figure}
	\includegraphics[width=\columnwidth]{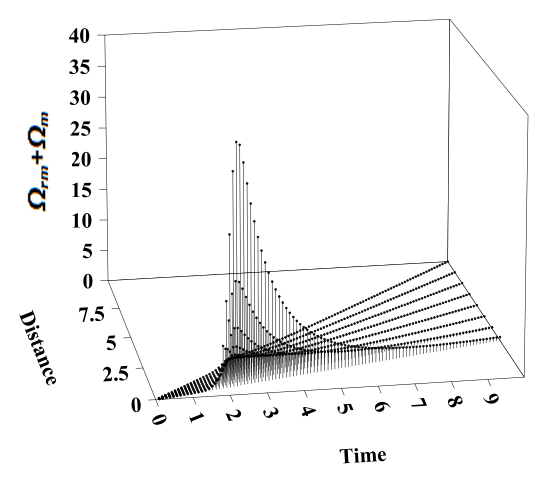}
    \caption{History of $\Omega_{rm}+\Omega_m$ the first moments after the BB allowing universe expansion at velocities from c to 0.3c using equation (\ref{matterdrag}) with $\textbf{k}$ of 0.01. Time metric can be from $\mu s$ to a few seconds after the BB. The values of $\Omega_{rm} + \Omega_m$ at lightspeed are presented as the points lying on the diagonal of the distance/time plane with the maximum normalised to $\Omega_{rm} + \Omega_m=1$ when $t=r$.}
    \label{fig:Figure2}
\end{figure}

In Figure \ref{fig:Figure2} we present values for $\Omega_{rm}$+$\Omega_m$ climbing toward a value normalised to $\approx1$, as the diagonally directed curve, as if the universe expanded at lightspeed. The situation where the universe expanded at lightspeed is probably unrealistic, so we present other solutions where the universe expansion occurred at fractions of lightspeed and note that $\Omega_{rm}$+$\Omega_m$ could have become much greater than 1 before \say{relaxing} towards much smaller values.

\begin{figure}
	\includegraphics[width=\columnwidth]{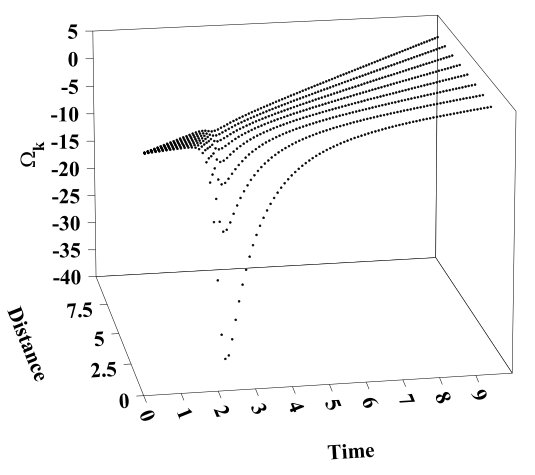}
    \caption{History of $\Omega_k$ the first moments after the BB allowing universe expansion at velocities from lightspeed to 0.3c using equation (\ref{matterdrag}) with $\textbf{k}$ of 0.01. Time from $\mu s$ to seconds after the BB. The values of $\Omega_k$ for expansion at c are presented as the points lying on the diagonal of the distance/time plane with the minimum at $\Omega_k=0$ when $t=r$.}
    \label{fig:Figure3}
\end{figure}

Figure 2 also illustrates that matter creation would have occurred rather quickly \textit{via} auto-catalysis. If $\Omega_{rm}$+$\Omega_m$ was $>1$ even for a short time, $\Omega_k$ would necessarily have been $<0$ and spacetime geometry became drastically elliptical. The effect on $\Omega_k$ is shown in Figure \ref{fig:Figure3} where minimum values are often less than 0. This is shown in this figure where points are plotted rather than curves drawn and where the expansion rates are significantly slower than lightspeed. Reversal towards the big crunch was avoided because the massive particles were then travelling at relativistic velocities or nearly so, hence $H$ was large. This is consistent with proposals for the mechanism whereby the Universe avoided gravitational collapse \cite{Barrow2018}. Nevertheless, a short period with $\Omega_{rm}$+$\Omega_m>1$, effecting a considerable drag on particle velocity, is consistent with a $H$ at recombination not too different than $H_0$. 

 The value for $\Omega_{rm}$+$\Omega_m$ in Figure \ref{fig:Figure2} decreases rapidly towards zero as matter was diluted in the expanding Universe. The small current value of $\Omega_m$ predicted here is consistent the results of BB calculations by \cite{Burles2001} and values when SNe Ia data ensembles are used for analysis \cite{Oeztas2008,Oeztas2014,Oeztas2015}. This also means that spacetime geometry fluctuated rapidly during particle creation. This wild change of universe geometry randomised particle and energy trajectories producing the homogeneity and isotropy we now observe. Also note that the smooth curve increase followed by the slow decline of $\Omega_{rm}$+$\Omega_m$ after the long period of particle creation means detection of the primordial gravitational wave(s) will be more difficult than if matter was created in an instantaneous manner.


One interesting result from our calculations is that with the dilution of matter during universe expansion at a rate close to that examined here, $\Omega_m$ eventually drops to less than 0.03 in agreement with studies of particle creation \cite{Burles2001,Burles2001a}. A second interesting finding is that our Universe exhibits quasi-Euclidean, gently elliptical and closed spacetime geometry as the inescapable result of particle dilution.

\section{Conclusion}
\label{conclusion}
The notion that $\Omega_k$ alone determines spacetime curvature and that $\Omega_k=0$ describes flat Universe geometry has long lead people astray. The historic assignments of 0 for flat geometry, $-K$ for elliptical and $+K$ for hyperbolic, appeals to the human sense of symmetry but is not useful in practice. Nowadays many astrophysicists report the current $\Omega_m\approx0.25$ to $0.30$ with $\Omega_{\Lambda}\approx0.70$ to $0.75$ and $\Omega_k\approx0$ describes a geometrically flat Universe. But how can flat geometry be consistent with a large value for $\Omega_m$ given that a large $\Omega_{\Lambda}$ is probably not true? Also, since our Universe can never be described with hyperbolic curvature it is unnecessary to place an $\Omega_k=0$ between elliptical and a hyperbolic geometry which can only exist in one's imagination. Such nagging problems have lead us to re-evaluate the nature of $\Omega_k$.

We have demonstrated for our early Universe that the key value for $\Omega_k$ is 1 which indicates Euclidean geometry and values approaching 1 describe quasi-Euclidean. We also suggest that $\Omega_k$ estimates not only the average large-scale geometric curvature but also the average distance between massive objects. That is, a value of $0<\Omega_k<1$ means more than elliptical curvature, it also means the average distance between matter ensembles, galaxies and galactic groups, is significant - which describes our current reality. 

Realisation that one is not forced to find $\Omega_k\approx0$ to explain a quasi-Euclidean universe geometry simplifies analysis and helps cosmologists understand heretofore puzzling observations and results. In a universe nearly devoid of matter, such as our current epoch, flat geometry is the expected limit. There is no obvious limit for $\Omega_m$ in smaller systems, which may locally be $\gg1$ meaning that $\Omega_k\ll0$ with little space between particles. When $\Omega_k$ is 0, this indicates $\Omega_{rm} + \Omega_m$ is equal to the critical matter density and the Friedmann approximation cannot be evaluated nor $D_L$ determined for this situation. When $\Omega_k<0$ the system suffers matter density greater than critical but must still exhibit elliptical curvature and little spacetime between objects. Much smaller worlds, for instance the spacetime of a black hole, may suffer $\Omega_k<0$ and certainly elliptical spacetime geometry.

This new explanation of $\Omega_k$ greatly simplifies many heretofore confusing results, for instance FLRW models evaluated using SNe Ia, $\gamma$-ray burst and HII/GEHR data \cite{Oeztas2015,Smith2020}.   Our analyses are also consistent with recent concerns that an accelerating Universe expansion and the $\Lambda$CDM model are not supported by hard data \cite{Sabulsky2019} or the searching analyses by others \cite{Dam2017,Ezquiaga2017}. One obvious test of a current $\Omega_k\approx1$  would be independent calculation of $\Omega_k$ for our Universe similar to that recently published by \cite{Valentino2020} but calculated without resorting to $\Omega_{\Lambda}$. Calculations using SNe and tip of the red giant branch data should be made allowing $\Omega_k$ and $H_0$ as free parameters with and without $\Omega_{\Lambda}$ to check our ideas. We think the results will help resolve the current \say{tension} about the Hubble constant value.

We describe the first moments of our Universe leading to a flat, homogeneous Universe without the need for a special event other than the hot BB. We have shown that strong fluctuations of spacetime geometry occurred a few seconds after the BB and may be the events responsible for our current condition. Our explanation and calculations of a simple autocatalytic model are consistent with the results of BB nucleosynthesis calculations. After particle creation, matter and energy densities were diluted as the Universe expanded, allowed $\Omega_k$ to approach 1 and spacetime to assume quasi-Euclidean geometry. On the other hand, a phase of inflationary expansion around the Planck epoch prior to the epoch of massive particle creation is not ruled out by anything presented here. Realising this allows discussion to return towards simpler versions of inflationary theory.





\bibliography{AstroRefsvers2}

\appendix
\section{Deceleration parameter}
The parameter $q$ is often used in cosmology describing the declining expansion rate of a universe and termed the \textit{deceleration parameter}. For reasons given above we are interested only in $\Omega_m$ and $\Omega_k$ and ignore $\Omega_{\Lambda}$ and $\Omega_r$ for this derivation. The derivation below is presented for students.

We first expand the Taylor series at $a$, the expansion factor, around any time $t^\prime$ which may be the current or most any time, past or future, as
  \begin{equation}
    a(t)=a(t^\prime)+\frac{da
    (t)}{dt}\Bigg\vert_{t^\prime}(t-t^\prime)+ \frac{1}{2}\frac{d^2a
    (t)}{dt^2}\Bigg\vert_{t^\prime}(t-t^\prime)^2+ \dots
  \end{equation}
To derive $q$ we divide the series by $a(t^\prime)$ and get following expression
  \begin{equation}
\frac{a(t)}{a(t^\prime)}=1+\frac{\dot{a}(t^\prime)}{a(t^\prime)}+\frac{1}{2}\frac{\ddot{a}(t^\prime)}{a(t^\prime)}(t-t^\prime)^2.
  \end{equation}
  This series is well known in cosmology as
 \begin{equation}
       \frac{a(t)}{a(t^\prime)}=1+H(t^\prime)(t-t^\prime)-
\frac{1}{2}q(t^\prime)H(t^\prime)^2(t-t^\prime)^2+\dots
  \end{equation}
  Here $H(t^\prime)={\frac{\dot a(t^\prime)}{a(t^\prime)}}$ is the Hubble-Lema\^{\i}tre
parameter and $q(t^\prime)$ the deceleration parameter at
  $t=t^\prime$ is
  \begin{equation} q(t^\prime)=-\frac{a(t^\prime) \ddot{a}(t^\prime)}{\dot{a}(t^{\prime})^2}=-\frac{1}{H(t^{\prime})^2}\frac{\ddot{a}(t^\prime)}{a(t^\prime)}\,,
  \label{deceleration}
  \end{equation}
  when $t^\prime$ is the current time we have
 \begin{equation} 
q(t_0)=-\frac{1}{H_0^2}\frac{\ddot{a}(t_0)}{a(t_0)}.
\label{deceleration2}
\end{equation} 
The negative sign in this equation implies that if \textit{a(t)} is positive (i.e. the universe expansion is accelerating) the deceleration parameter must be negative.
We will now relate $q(t_0)$ as $q_0$ to the Friedman equations without $\Lambda$, but with Newton's G and $\rho$ matter density and $p$ pressure; relations \ref{Fried1} and \ref{Fried2}, as the right-hand side of equation (\ref{decelratio}) below
\begin{equation}
q_0=-\left(\frac{a}{\dot{a}}\right)^2\frac{\ddot{a}}{a}=-\Bigg[\frac{-\frac{4 \pi G}{3}\left(\rho+\frac{3p}{c^2}\right)}{\frac{8\pi G}{3}\rho - \frac{Kc^2}{a^2}}\Bigg].
\label{decelratio}
\end{equation}
This relationship allows us to eventually express $q_0$ in terms of $H$, $\Omega_m$ and $\Omega_k$. For the current epoch we can ignore the radiation pressure as being minuscule, we have the slightly simplified
\begin{equation}
q_0=-\left(\frac{a}{\dot{a}}\right)^2\frac{\ddot{a}}{a}=-\Bigg[\frac{-\frac{4 \pi G\rho}{3}}{\frac{8\pi G}{3}\rho - \frac{Kc^2}{a^2}}\Bigg].
\label{decelratio2}
\end{equation}
We can further simplify this to
\begin{equation}
\frac{1}{q_0}=-\Bigg[\frac{\frac{8\pi G}{3}\rho - \frac{Kc^2}{a^2}}{-\frac{4 \pi G\rho}{3}}\Bigg]=2-\frac{{\frac{Kc^2}{a^2}}}{{\frac{4 \pi G\rho}{3}}}.
\label{decelratio3}
\end{equation}
We now rewrite this relationship in terms of $\Omega_k$, $\Omega_m$ and $H^2$ as
\begin{equation}
\frac{1}{q_0}=2+\frac{\Omega_k H^2}{\frac{\Omega_m H^2}{2}}=2+\frac{\Omega_k}{\frac{\Omega_m}{2}}\,,
\label{decelratio4}
\end{equation}
or as a single term for $q_0$ we have
\begin{equation}
\frac{1}{q_0}=\frac{2(\Omega_m+\Omega_k)}{\Omega_m}.
\label{simpledecel}
\end{equation}
 We remember that $\Omega_m+\Omega_k=1$ without much influence from $\Omega_r$, we have
\begin{equation}
q_0=\frac{\Omega_m}{2}.
\end{equation}
So deceleration is independent of the value for $\Omega_k$, whether one believes a flat universe is described by $\Omega_k=0$ or $1$.

In the absence of other parameters, with $\Omega_m+\Omega_k=1$ we arrive at
\begin{equation}
q(t_0)=\frac{\Omega_m}{2}.
\label{simpledece2}
\end{equation}
The deceleration parameter value is only dependent on $\Omega_m$ and consistent with a universe containing matter with the current geometry described as elliptical.


\section{Light and the early Universe}
While equations (\ref{supernova}) and (\ref{earuni}) are useful for modelling the current epoch we can test these as $\Omega_m$ and $\Omega_r$ approach zero with $\Omega_k$ approaching 1, to get an idea of the properties of light travelling in the early Universe near the instant of massive particle creation. 

\begin{figure}
	\includegraphics[width=\columnwidth]{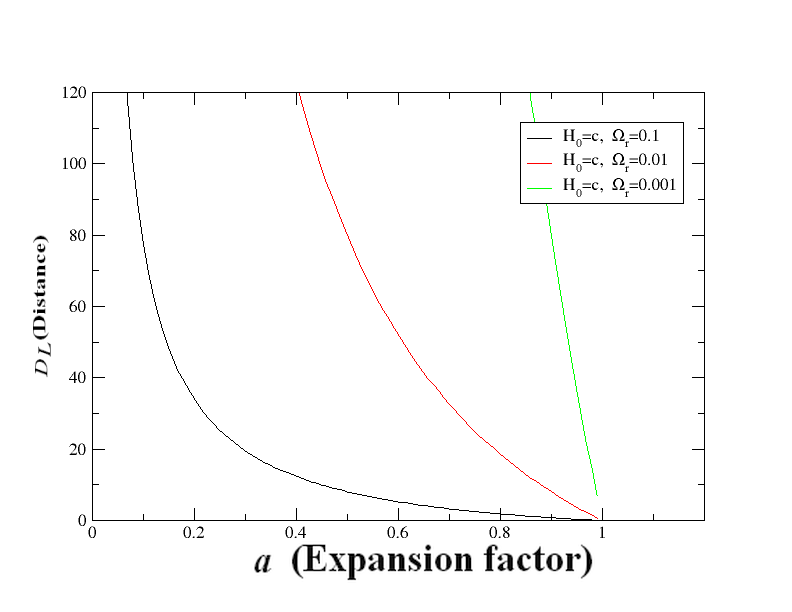}
    \caption{$D_L$ \textit{vs.} \textit{a} with $H_0=c=1$ for a 
    universe with $\Omega_r$ from 0.1 to 0.001, left to right, with corresponding
    values of $\Omega_k$, calculated with equation (\ref{earuni}).}
    \label{fig:Fig2}
\end{figure}
Again as $D_L$ approaches extreme dependence on the expansion factor with decreasing high energy density and as \textit{a} approaches 1 most any distance is allowed as displayed in Figure \ref{fig:Fig2}. An $\Omega_k$ of 1 is consistent with a universe devoid of gravity with radiation everywhere in a spacetime without bounds.

\section{Graph SNe Ia data}
Data including standard errors calculated from the first set of 37 SNe Ia emissions presenting claims for an accelerating Universe expansion \cite{Riess1998}. Location of our earth at (1,0) is bottom right is used as data pair number 38 which is the one data pair without error. Note the very large errors of distant emissions which were damped in the \textit{mag vs.} z analysis by the High-Z Supernova Search Team thus negating the validity of their analysis.
\begin{figure}
	\includegraphics[width=\columnwidth]{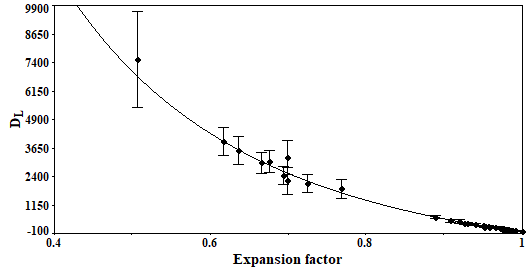}
    \caption{$D_L$ (as Mpc) \textit{vs.} the expansion factor, parameters from this curve are listed in Table 1 as the $\Omega_m+\Omega_k=1$ model.}
    \label{fig:DistanceData}
\end{figure}

\end{document}